\documentclass[12pt,epsf,amssymb,qsymbols]{article}
\usepackage{tabularx}
\usepackage{array}
\usepackage{graphics}
\usepackage{graphicx}
\usepackage{epsfig}
\usepackage{amsmath}
\usepackage{amssymb}
\usepackage{ulem}
\usepackage{figlatex}
\usepackage{rotating}
\makeatletter

\usepackage{verbatim}

\newcommand{\bea}{\begin{eqnarray}}
\newcommand{\eea}{\end{eqnarray}}
\newcommand{\beq}{\begin{equation}}
\newcommand{\eeq}{\end{equation}}
\newcommand{\gev}{{\rm GeV}}
\newcommand{\mev}{{\rm MeV}}
\newcommand{\kev}{{\rm keV}}

\newcommand{\pdir}{p\kern -5.2pt\raise 0.2ex\hbox {/}}
\newcommand{\vdir}{v\kern -5.75pt\raise 0.15ex\hbox {/}}
\newcommand{\kdir}{k\kern -5.75pt\raise 0.15ex\hbox {/}}
\newcommand{\epsdir}{\epsilon\kern -5.0pt\raise 0.15ex\hbox {/}}
\newcommand{\bvdir}{\bar{v}\kern -5.75pt\raise 0.15ex\hbox {/}}
\newcommand{\Ddir}{D\kern -7.75pt\raise 0.20ex\hbox {/}}
\newcommand{\Adir}{A\kern -7.75pt\raise 0.20ex\hbox {/}}
\newcommand{\ldir}{l\kern -5.0pt\raise 0.2ex\hbox{/}}
\newcommand{\varepsdir}{\varepsilon\kern -5.5pt\raise 0.15ex\hbox{/}}

\newcommand{\nn}{\nonumber}
\makeatother

\begin{document}
\thispagestyle{empty} 
\begin{flushright}
\begin{tabular}{l}
 \\
\end{tabular}
\end{flushright}
\begin{center}
\vskip 3.0cm\par
{\par\centering \textbf{\LARGE  
\Large \bf $D^\ast \to D\pi$ and $D^\ast \to D\gamma$ decays:\\
\vspace*{4mm}
Axial coupling and Magnetic moment of $D^\ast$ meson }}\\
\vskip 1.25cm\par
{\scalebox{.9}{\par\centering \large  
\sc Damir Be\'cirevi\'c and Benjamin Haas}}
{\par\centering \vskip 0.75 cm\par}
{\sl 
Laboratoire de Physique Th\'eorique (B\^at.~210)~\footnote{Laboratoire de Physique Th\'eorique est une unit\'e mixte de recherche du CNRS, UMR 8627.}\\
Universit\'e Paris Sud, Centre d'Orsay,\\ 
F-91405 Orsay-Cedex, France.}\\
\vskip1.cm
 
{\vskip 0.35cm\par}
\end{center}

\vskip 0.55cm
\begin{abstract}
The axial coupling and the magnetic moment of   $D^\ast$-meson or, more specifically, the couplings $g_{D^\ast D\pi}$ and $g_{D^\ast D\gamma }$, encode the non-perturbative 
QCD effects describing the decays $D^\ast \to D\pi$ and $D^\ast \to D\gamma$. We compute these quantities by means of lattice QCD with $N_{\rm f}=2$ dynamical quarks, by employing the 
 Wilson  ({\sl ``clover"}) action. On our finer lattice ($a\approx 0.065$~fm) we obtain: $g_{D^\ast D\pi^+}=20\pm 2$, and $g_{D^{\ast 0} D^0\gamma }=2.0\pm 0.6~\gev^{-1}$. This is the first 
determination of  $g_{D^{\ast 0} D^0\gamma }$ on the lattice. We also provide a short phenomenological discussion and the comparison of our result with experiment and with the  
results quoted in the literature. 
\end{abstract}
\vskip 2.2 cm 
\setcounter{page}{1}
\setcounter{footnote}{0}
\setcounter{equation}{0}
\noindent

\renewcommand{\thefootnote}{\arabic{footnote}}

\newpage
\setcounter{footnote}{0}
\section{Introduction}
Measuring the width of the nearest resonances in the charmed meson spectrum is a challenging experimental task. The methods developed by the CLEO-collaboration allowed 
them to make the first measurement of the charged vector meson width, $\Gamma(D^{\ast +})=96\pm 22$~keV~\cite{cleo}. Unfortunately 
no similar attempt has been made in  the experiments at $B$-factories or at CLEO-c. 
The experimental value for  $\Gamma(D^{\ast +})$ turned out to be very interesting because it provided us with the quantity allowing to check on various theoretical 
tools that are being used to compute the phenomenologically interesting hadronic matrix elements such as those needed for the decay constants, form factors, 
bag parameters and so on. The coupling $g_c$, that can be extracted from $\Gamma(D^{\ast +})$, as discussed in later sections of this paper, 
appeared to be much larger than predicted by many quark models and by all the techniques of QCD sum rules (see ref.~\cite{alain} for a discussion).  
The CLEO result was however consistent with the old fashioned Adler-Weisberger sum rule combined with phenomenological observations made with baryons~\cite{alain}. 

The first lattice calculation of this coupling in the charmed sector was made in ref.~\cite{Abada:2002xe}, where it was possible to extract this coupling ($g_c$) in the soft pion limit 
without running into notorious difficulties of dealing with the final state interactions in non-leptonic decays on the euclidean lattice. In this paper we provide an update to that result. 
We use the same action and the same methodology as in ref.~\cite{Abada:2002xe} but the major qualitative difference with respect to  ref.~\cite{Abada:2002xe}  is that here  
we get rid of the quenched approximation. The gauge field configurations used in this work contain the fluctuations of the $N_{\rm f}=2$ mass-degenerate dynamical light quarks. 
Besides we also implement several technical improvements to make the extraction of the form factors from the correlation functions computed on the lattice cleaner (double ratios of 
correlation functions, twisted boundary conditions). 
On the basis of our results we conclude that the lattice evaluation of $D^\ast \to D\pi$ decay is consistent with experiment, and points towards the upper end of  the measured $\Gamma(D^{\ast +})$~\cite{cleo}.

Beside that quantity, in this paper we report on the first lattice determination of the soft photon coupling to the lowest $D$-meson states, $g_{D^\ast D\gamma}$. 
In particular we obtain $\Gamma(D^{\ast 0}\to D^0\gamma) = 25\pm 13$~keV, in  good agreement with experiment, although our error bars are still too large to make any stronger statement. 
The two quantities computed here are obviously phenomenologically interesting since the experimental information about these decays already exists. Therefore they can be used to check on  
various quark model calculations and to perhaps refine the QCD sum rule analyses. Moreover the quantities computed here are particularly interesting to the lattice practitioners. $g_c$-coupling 
is a parameter which appears in the expressions obtained in heavy meson chiral perturbation theory (HMChPT)  
that are being used to guide the chiral extrapolations of the phenomenologically interesting quantities computed on the lattice, such as the $D$-meson decay constant and its semileptonic decay form factors.
Instead, $g_{D^\ast D\gamma}$ is important in controlling the impact of the structure dependent term in  the radiative leptonic $D$-meson decays in the region of phase space in which  the photon is soft.  
Finally, the radiative decay $D^{\ast }\to D\gamma$ can be used as a benchmark calculation to compare the lattice approaches among themselves and with experiment. The obvious advantage 
is that in this case one goes beyond the hadronic spectrum and computes the hadronic matrix element but there is  no need for a CKM parameter to make the precision test of lattice QCD vs. experiment. 

The present paper is organized as follows: we first define the hadronic couplings which are the subject of this paper, and relate them to the form factors that are to be computed non-perturbatively;
in Sec.~3 we explain the strategy to compute the relevant hadronic matrix elements on the lattice, the strategy which we then implement   in Sec.~4 where we also  present the results directly accessible from our lattices; in Sec.~5 we 
 provide a phenomenological discussion and compare our results with experiment and with various model calculations; we finally conclude in Sec.6. 
 
\section{Definitions}
\setcounter{equation}{0}
In this section we define the couplings $g_{D^\ast D\pi}$ and $g_{D^\ast D\gamma}$, and relate them to the matrix elements that can be computed on the lattice.
\subsection{$g_{D^\ast D\pi}$}
Generically, the coupling of the pseudoscalar ($P$) and vector ($V$)  mesons to a soft pion is defined as 
\bea
\langle P(k) \pi(q) \vert V(p,\lambda)\rangle = (e_\lambda \cdot q)\ g_{VP\pi} \,,
\eea
where $q=p-k$ is the pion momentum and $\lambda$ labels the polarization state of the vector meson.  Physically this matrix element describes the amplitude of the soft pion emission process,  
$V\to P\pi$,  such as $K^\ast \to K\pi$ and $D^\ast \to D\pi$, or the kinematically forbidden, but nonetheless very interesting processes, $B^\ast\to B\pi$ and $\omega \to \rho\pi$.   
In this paper we will focus onto $g_{D^\ast D\pi}$,  and follow a rather standard procedure, first proposed in ref.~\cite{deDivitiis:1998kj} and then implemented in refs.~\cite{Abada:2002xe,deDivitiis:1998kj,static}. 
To describe it in just a few lines we first define the matrix element of the axial current $A_\mu = \bar q \gamma_\mu\gamma_5 q$ (here ``$q$" denotes the light quark, $u$ or $d$) 
\bea\label{ffsA}
\langle D(k)\vert A^\mu \vert D^\ast (p,\lambda)\rangle = 2m_VA_0(q^2)\frac{\epsilon_\lambda\cdot q}{q^2}q^\mu+(m_D+m_{D^\ast})A_1(q^2)\left(\epsilon_\lambda^\mu-\frac{\epsilon_\lambda\cdot q}{q^2}q^\mu\right)\nonumber\\
+A_2(q^2)\frac{\epsilon_\lambda\cdot q}{m_D+m_{D^\ast}}\left( p^\mu+k^\mu-\frac{m_{D^\ast}^2-m_{D}^2}{q^2}q^\mu\right) \,,
\eea
conveniently parameterized in terms of three form factors, $A_{0,1,2}(q^2)$.  We restrain our discussion  to the charged pion case, so that no anomalous term appears in the axial current.  
At $q^2$ close to $m_\pi^2$ we can use the reduction formula and write
\bea
{f_\pi  m_\pi^2\over m_\pi^2 - q^2} \langle D(k) \pi(q) \vert D^\ast (p,\lambda)\rangle = \langle D(k)\vert \partial_\mu A^\mu \vert D^\ast(p,\lambda)\rangle\, ,
\eea
which at $q^2=0$ leads to 
\bea
g_{D^\ast D\pi} = {2 m_{D^\ast}\over f_\pi} A_0(0)\,.
\eea
This formula is not useful for lattice QCD because the form factor $A_0(q^2)$ is dominated by the pion ($J^P=0^-$-state in the $t$-channel), so that both the $q^2$ dependence of $A_0(q^2)$ and 
the mass dependence of $A_0(0)$ are extremely steep.  However  from the fact that no massless state can couple to the axial current, we can benefit from the condition that 
\bea
2m_{D^\ast} A_0(0)-(m_{D^\ast}+m_D)A_1(0)-\left(m_{D^\ast}-m_D\right) A_2(0) = 0\,,
\eea
and eventually arrive at
\bea\label{defA}
g_{D^\ast D\pi}= \frac{m_{D^\ast}+m_D}{f_\pi} A_1(0)\left[1+{m_{D^\ast}-m_D\over m_{D^\ast}+m_D}{A_2(0)\over A_1(0)}\right]  \,.
\eea
In other words  the problem of computing the coupling $g_{D^\ast D\pi}$ is reduced  to the problem of computing the form factor $A_1(0)$ and the ratio $A_2(0)/A_1(0)$.
We stress again that  in the above definitions the soft pion is charged $g_{D^\ast D\pi} \equiv g_{D^\ast D\pi^+}$ . The coupling to the neutral pion is 
related to the one we compute here via isospin, i.e., $ g_{D^\ast D\pi^+}^2 = 2  g_{D^\ast D\pi^0}^2$.

\subsection{$g_{D^\ast D\gamma}$}

The coupling of the pseudoscalar ($P$) and vector ($V$)  mesons to a photon is defined as
\bea
\langle \gamma(q,\eta_{\lambda'})P(k)\vert V(p,\epsilon_\lambda)\rangle  =e\ \varepsilon_{\mu\nu\alpha\beta} \  
\eta_{\lambda'}^\mu{\epsilon_{\lambda}}^{\nu} p^\alpha k^\beta  \ g_{VP\gamma}\,,
\eea
which we will accede by computing the matrix element of the electromagnetic current, namely
\bea\label{def-vectorFF}
\langle P(k) \vert  J^{\rm em}_\mu \vert V(p,\epsilon_\lambda) \rangle = 
e\ \varepsilon_{\mu\nu\alpha\beta}\  \epsilon_\lambda^\nu p^\alpha k^\beta \ \frac{2\ V(q^2)}{ m_V+m_P }\,,
\eea
where $J^{\rm em}_\mu ={\cal Q}_Q \bar Q \gamma_\mu Q + {\cal Q}_{\bar q} \bar q \gamma_\mu q$, and the matrix element is expressed in a rather standard way, 
i.e. in terms of the form factor  $V(q^2)$. In the above expressions $e=\sqrt{4\pi \alpha_{\rm em}}$, and ${\cal Q}_{Q,\bar q}$ is the charge of the heavy quark and the light antiquark (or vice versa). 
It is very important to keep track of the relative sign difference between the two terms in $J^{\rm em}_\mu$. In the case of $D$-mesons we have
\bea\label{defV}
&&g_{D^{\ast +} D^+ \gamma} ={ 2\over m_D + m_{D^\ast} } F_d(0)\,,\,\, {\rm with}\quad F_d(0) ={2\over 3} V^{qq}(0)   \left[ -{1\over 2} + {V^{cc}(0)\over V^{qq}(0)} \right]  \ ,\nn\\
&&g_{D^{\ast 0}D^0 \gamma} ={ 2\over m_D + m_{D^\ast} } F_u(0)\,,\,\, {\rm with}\quad F_u(0) ={2\over 3} V^{qq}(0)   \left[ 1 + {V^{cc}(0)\over V^{qq}(0)} \right]  \  .
\eea
We evidently separated the  `charm' ($\bar c  \gamma_\mu c$) from the `up/down' ($\bar q  \gamma_\mu q$) contributions to 
the electromagnetic current the matrix elements of which will be computed separately.  We should emphasize that we do  not consider the isospin violating effects and we consistently take $m_u=m_d\equiv m_q$.
Notice that, contrary to the  pionic coupling ($g_{D^\ast D\pi}$), the coupling to the soft photon ($g_{D^\ast D\gamma}$) is dimension-full,  and it is a measure of the magnetic moment 
of the $D^\ast$-meson.  To make the lattice computation more straightforward our target will be to compute the dimensionless form factor  $F_{u,d}(0)$.

\section{Correlation functions computed on the lattice}

In this section we list the correlation functions that are to be computed on the lattice, define the convenient double ratios of correlation functions which lead to the desired 
form factors. We then discuss the method that helps  exploring the kinematical configurations in which when $\vec q\neq 0$.

\subsection{ Axial form factors }
In order to reach the matrix element~(\ref{ffsA}) we need to compute the following three-point correlation functions
\bea\label{def-1}
C^{(1)}_{\mu\nu}(\vec q;t)&=& \sum_{\vec x,\vec y } \langle {\cal V}_\mu(\vec  0,0) A_\nu (\vec x, t)   {\cal P}^\dag (\vec y, t_S) e^{-i\vec q\vec x}\rangle\,, \nn \\
C^{(2)}_{\mu\nu}(\vec q;t)&=& \sum_{\vec x,\vec y } \langle {\cal P} (\vec  0,0) A_\mu (\vec x, t)   {\cal V}_\nu^\dag (\vec y, t_S) e^{-i\vec q\vec x}\rangle\,, \nn \\
E^{(1)}(\vec q;t)&=& \sum_{\vec x,\vec y } \langle {\cal P}(\vec 0,0 ) V^q_0 (\vec x,t)   {\cal P}^\dag (\vec y, t_S) e^{-i\vec q\vec y} \rangle\,, \nn \\
E^{(2)}(t)&=& {1\over 3} \sum_{i,\vec x,\vec y } \langle {\cal V}_i(\vec 0,0 ) V^q_0 (\vec x,t)   {\cal V}^\dag_{i} (\vec y, t_S) \rangle\,, \nn
\eea
where ${\cal V}_\mu = \bar c \gamma_\mu q$, and  ${\cal P} = \bar c \gamma_5 q$ are the interpolating operators for the vector and the pseudoscalar 
meson respectively, while 
$V^{q}_\mu= \bar q \gamma_\mu q$ and $A_\mu= \bar q \gamma_\mu\gamma_5 q$.  The sources are fixed at $t=0$ and $t_S=24$, and  the matrix element is extracted from the usual time dependence 
of the correlation functions.  In recent years it became clear that a more reliable information can be extracted if one cancels the exponential time dependencies 
in the correlation functions by combining them into suitable double ratios~\cite{BB,DD,KK}. 
This is why in eq.~(\ref{def-1}) we also defined the elastic correlation functions, $E^{(1,2)}$. 
In the case in which both $D$- and $D^\ast$-mesons are at rest the only contributing form factor  is $A_1(q^2_{\rm max})$, with $q^2_{\rm max}=(m_{D^\ast} - m_D)^2$. In that case, 
 the convenient double ratio to consider is
\bea\label{defR0}
R_0(t) = { C^{(1)}_{i 0}(\vec 0;t)C^{(2)}_{0 i}(\vec 0;t) \over E^{(1)}(\vec 0;t)E^{(2)}(\vec 0;t)} &\to& 
{ \langle D(\vec 0)\vert A_0\vert  D^\ast (\vec 0) \rangle \langle D^\ast (\vec 0)\vert A_0\vert  D(\vec 0) \rangle \over 
 \langle D(\vec 0)\vert V_0\vert  D (\vec 0) \rangle  \langle D^\ast (\vec 0)\vert V_0\vert  D^\ast (\vec 0) \rangle } \nn\\
&& \nn\\
 &=& {\left[ (m_{D^\ast} + m_D) A_1(q^2_{\rm max})\right]^2\over 4 m_D m_{D^\ast}}.
\eea
Since in eq.~(\ref{defA}) we need the form factors at $q^2=0$, an extra step is needed. 
As it is well known, on the periodic cubic lattice of size $L$, the smallest momentum that can be given to a particle is $q_{\rm min} = (2\pi/L)$, 
which on the currently accessible  lattices would be far too large, and would eventually push $q^2$'s to relatively large negative values. It has been shown in refs.~\cite{nazario} 
that by imposing the twisted boundary conditions on one of the quark propagators in the correlation function, one can explore the arbitrary small momenta of  the hadron associated to that ``twisted" quark. 
A simple recipe to implement this idea in practical calculations consists in rephasing the gauge links as 
\bea\label{twisted}
U_\mu(x) \to U^\theta_\mu(x) = e^{i \theta_\mu/L}U_\mu(x),\quad {\rm where }\; \theta_\mu = (0,\vec \theta) \,,
\eea
and then inverting the propagator on the lattice with periodic boundary conditions, i.e. by inverting the Wilson-Dirac operator on such a rephased gauge field configuration, 
$S_q(x,0;U^\theta) \equiv \langle q(x)\bar q(0)\rangle$. Finally the twisted propagator is obtained as
\bea
S_q^{ \vec \theta} (x,0;U) = e^{i \vec \theta\cdot \vec x/L} \ S_q(x,0;U^\theta) \,.
\eea
The net effect is that the resulting ground state  extracted at large time separations between the sources in  two-point correlation functions, e.g. 
\bea
 \sum_{\vec x } \langle {\rm Tr}\left[ S_q^{ \vec \theta} (0,x;U) \gamma_5 S_q (x,0;U)  \gamma_5  \right] \rangle  \longrightarrow 
 {1
 \over 2 E_\pi} \vert \langle 0\vert \bar q \gamma_5 q\vert  \pi \rangle\vert^2 \  e^{- E_\pi t} \,,
\eea
satisfies the following dispersion relation~\cite{nazario,diego} 
\bea
E_\pi^2 = m_\pi^2 +   {\vert \vec \theta\ \! \vert^2 \over L^2}\;.
\eea
Therefore we can give as small a momentum to a daughter pseudoscalar meson in eqs.~(\ref{ffsA},\ref{def-vectorFF})  as we want. An important discussion and description of the same 
physics by means of a low-energy effective theory in ref.~\cite{chris-giovanni} assert that adding a small twisted angle to the boundary condition on the valence quark, but not on the sea quarks, 
has a negligible impact on the low energy QCD observables, such as $f_{K,\pi}$ and  $m_{K,\pi}$, and we shall neglected it  in what follows. 
For further reference, when using the twisted quark propagator, we take vector  $\vec \theta$ to be of the form $\vec \theta = (\theta_0, \theta_0,\theta_0)$. We will also always 
keep the decaying vector meson at rest.   More specifically,  we compute the correlation 
functions~(\ref{def-1}) as follows 
\bea
C^{(1)}_{\mu\nu}(\vec q;t)&=&  \langle \sum_{\vec x,\vec y}{\rm Tr}\left[ S_c(y,0;U) \gamma_\mu  S_q (0,x;U)\gamma_\nu \gamma_5 S_q^{ \vec \theta}(x,y;U)  \gamma_5 \right]\rangle \,, \nn \\
C^{(2)}_{\mu\nu}(\vec q;t)&=&  \langle \sum_{\vec x,\vec y}{\rm Tr}\left[  S_c(y,0;U) \gamma_5 S^{ \vec \theta}_q (0,x;U)\gamma_\mu \gamma_5 S_q(x,y;U) \gamma_\nu \right]\rangle \,, \nn \\
E^{(1)} (\vec q;t)&=&  \langle \sum_{\vec x,\vec y}{\rm Tr}\left[  S_c(y,0;U) \gamma_5 S_q^{ \vec \theta} (0,x;U)\gamma_0  S_q^{ \vec \theta}(x,y;U)\gamma_5\right]\rangle \,, \nn \\
E^{(2)} (t)&=& {1\over 3} \langle \sum_{i,\vec x,\vec y}{\rm Tr}\left[  S_c(y,0;U) \gamma_i S_q (0,x;U)\gamma_0  S_q(x,y;U)\gamma_i\right]\rangle\,,
\eea
where $\vec q=\vec \theta/L$, and  $\langle\dots\rangle$ denotes the average over  gauge field configurations, $U$. To extract the form factor at $|\vec q \ |\neq 0$, we first note that for large time separations one has 
\bea
&&\hspace*{-21mm}\widetilde C^{(2)}_{i j}(\vec q;t) \equiv {1\over 3}  \sum_{i=1}^3 C^{(2)}_{i i}(\vec q;t) - {1\over 6}  \sum_{i,j=1}^3 \left.C^{(2)}_{i j}(\vec q;t)\right|_{i\neq j} \longrightarrow \hfill \cr
&& {\langle D(\vec q)\vert {\cal P}\vert 0\rangle \over 2 E_D} e^{- E_Dt} \times (m_D + m_{D^\ast}) A_1(q^2) \times  
{\langle 0 \vert {\cal V}_i\vert D^\ast(\vec 0) \rangle \over 2 m_{D^\ast} } e^{- m_{D^\ast} (t_S-t)}\,,
\eea
and therefore the convenient double ratio to extract $A_1(q^2)$ looks very similar to $R_0(t)$ in eq.~(\ref{defR0}), and reads
\bea\label{defR1}
R_1(t) = { \widetilde C^{(1)}_{ji}(\vec q;t) \widetilde C^{(2)}_{i j}(\vec q;t) \over E^{(1)}(\vec q;t)E^{(2)}(\vec 0;t)} \to  {\left[ (m_{D^\ast} + m_D) A_1(q^2)\right]^2\over 4 E_D m_{D^\ast}}\,.
\eea
With our $D^\ast$-meson at rest the point $q^2=0$ is reached when the twisting angle 
\bea
\theta_0^\prime = {L\over \sqrt{3} } { m_{D^\ast}^2 - m_{D}^2\over 2 m_{D^\ast} }\,.
\eea
In practice, we give a few values of $\theta_0$ around $\theta_0^\prime$ so that we can interpolate the form factors to $q^2=0$. As for the correcting factor $A_2(q^2)/A_1(q^2)$ we extract it from the following 
ratio of the correlation functions 
\bea\label{defR2}
R_2(t)= -{   C^{(2)}_{0i}(\vec q;t) + \displaystyle{\frac{E_D-m_{D^\ast}}{ q_i}}  \widetilde C^{(2)}_{i j}(\vec q;t)  \over  \widetilde C^{(2)}_{i j}(\vec q;t)  } \to  {2 q_i m_{D^\ast}\over 
 (m_{D^\ast} + m_D)^2 } \ {A_2(q^2)\over A_1(q^2)}\,,
\eea
where, with the definition~(\ref{twisted}), $q_i=\theta_0/L$. This concludes our strategy for computing the axial coupling, $g_{D^\ast D\pi}$, on the lattice. 

\begin{figure}[t!]
\begin{center}
\epsfig{file=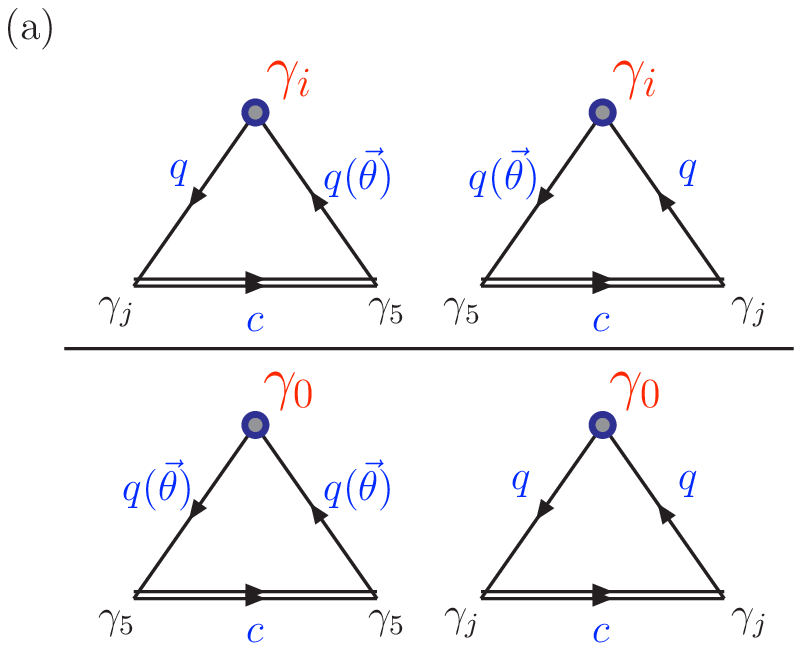, height=7.cm}\hspace{18mm}
\epsfig{file=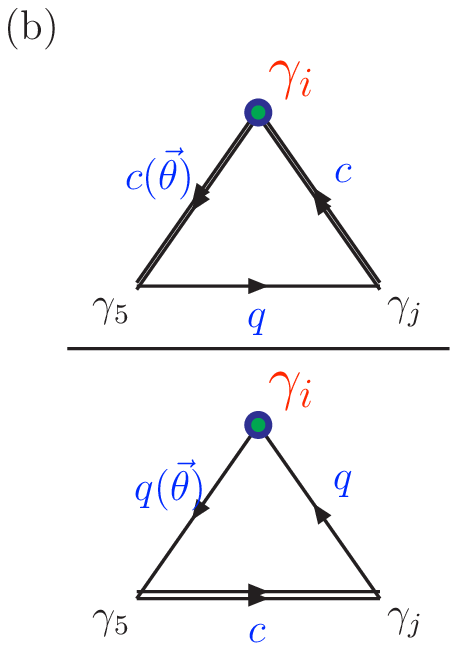, height=7.cm}
\caption{\label{figA}\footnotesize{\sl 
The Feynman diagrams used to compute the ratios $R^V_1(t )$ and $R^V_2(t )$ in eqs(\ref{RV1}) and (\ref{RV2}) respectively. The full dot symbols are used to denote the transition operator, the matrix element of which we are interested in. The symbol ``$\vec \theta$" with a given quark line indicates that the twisted boundary conditions were imposed in the propagator inversion of that particular quark. The single- and double-line distinguish between 
the light and the heavy quark propagator.}} 
\end{center}
\end{figure}
\subsection{ Vector form factors }
Concerning the coupling to a soft photon, we should first emphasize that our vector meson is at rest so that the matrix element~(\ref{def-vectorFF}) simply becomes 
\bea
\langle D(\vec k) \vert  J_{\rm em}^0 \vert D^{\ast}(\vec 0,\epsilon^\lambda) \rangle = 0\,,
\quad {\rm and}\quad
\langle D(\vec k) \vert  \vec J_{\rm em} \vert D^\ast (\vec 0,\epsilon^\lambda) \rangle 
=\   ( \vec \epsilon_\lambda \times \vec k)\  \frac{2e \ m_{D^\ast} }{ m_{D^\ast}+m_D }\ V(q^2)\,.
\eea
The correlation functions needed in this case are 
\bea\label{corrVV}
C^{(1qq)}_{ij}(\vec q;t)&=&  \langle \sum_{\vec x,\vec y}{\rm Tr}\left[ S_c(y,0;U) \gamma_i  S_q (0,x;U)\gamma_j  S_q^{ \vec \theta}(x,y;U)  \gamma_5 \right]\rangle \,, \nn \\
C^{(1cc)}_{ij}(\vec q;t)&=& - \langle \sum_{\vec x,\vec y}{\rm Tr}\left[  S_c(x,0;U) \gamma_i S_q (0,y;U)\gamma_5 S_c^{ \vec \theta}(y,x;U) \gamma_j \right]\rangle \,, \nn \\
C^{(2qq)}_{ij}(\vec q;t)&=&  \langle \sum_{\vec x,\vec y}{\rm Tr}\left[ S_c(y,0;U) \gamma_5  S_q^{ \vec \theta} (0,x;U)\gamma_i  S_q(x,y;U)  \gamma_j \right]\rangle \,.
\eea
Notice in particular the sign difference  of  ``$\bar c\vec \gamma c$"- with respect to the ``$\bar q\vec \gamma q$"-part of the electromagnetic current, which comes from the Wick contraction. 
Physically that accounts for the spin 
difference between the $D$ and $D^\ast$ states. Finally, that minus-sign compensates the relative sign difference between the electric charge of the  light and heavy quark/antiquark in the heavy-light meson. 
That sign-flip has been properly taken into account when writing the expressions~(\ref{defV}).  Finally from the double ratio
\bea\label{RV1}
R^V_1(t ) =-  {  {\cal C}^{(1qq)}_{ji}(\vec q;t) {\cal C}^{(2qq)}_{i j}(\vec q;t) \over E^{(1)}(\vec q;t)E^{(2)}(\vec 0;t)}  \to  { |{\vec q }|^2 \over 3}{m_{D^\ast} \over  E_D}  {[V^{qq}(q^2)]^2 \over (m_{D^\ast} + m_D)^2 }  \,,
\eea
we extract the desired form factor. In the above expression we use the symmetry of the problem and average over the equivalent indices, e.g.
\bea
 {\cal C}^{(2qq)}_{ij}(\vec q;t) &=& {1\over 6} \left[  C^{(2qq)}_{12}(\vec q;t) + C^{(2qq)}_{23}(\vec q;t) +C^{(2qq)}_{31}(\vec q;t)\right. \cr
 &&\left. \,\, -C^{(2qq)}_{21}(\vec q;t) -C^{(2qq)}_{32}(\vec q;t) -C^{(2qq)}_{13}(\vec q;t) \right]\,.
\eea
Notice yet another ``$-$"  sign in eq.~(\ref{RV1}) which is meant to compensate the sign difference between the two matrix elements accessed from ${\cal C}^{(2qq)}_{ij}(\vec q;t) $ (photon emission, $D^\ast \to D\gamma$) and 
from   ${\cal C}^{(1qq)}_{ij}(\vec q;t) $ (photon absorption, $\gamma D\to D^\ast$).  
Finally, the ratio of the two form factors is computed through
\bea\label{RV2}
R^V_2(t ) =-  {  {\cal C}^{(2cc)}_{ij}(\vec q;t)\over  {\cal C}^{(2qq)}_{i j}(\vec q;t)}  \to   {V^{cc}(q^2) \over V^{qq}(q^2) } \,, 
\eea 
where $ {\cal C}^{(2cc)}_{ij}(\vec q;t)$ is of the same form as  $ {\cal C}^{(2qq)}_{i j}(\vec q;t)$ after replacing $q\leftrightarrow c$ in eq.~(\ref{corrVV}). 
The Feynman diagrams of the double ratios leading to $R^V_1(t )$ and  $R^V_2(t )$ are shown in fig.~\ref{figA}a and \ref{figA}b respectively.

\subsection{Improvement and renormalization}
The gauge field configurations that are being used in this work are obtained with the Wilson plaquette gauge action, while the effects of dynamical quarks are incorporated by using the
${\cal O}(a)$ non-perturbatively improved  Wilson fermions, with ``$a$" being the lattice spacing. We consider only the fully unquenched situations in which the light valence 
and the sea quarks are mass degenerate.  To fully respect the ${\cal O}(a)$ improvement we also need to improve the operators. The subtraction of ${\cal O}(a)$ effects needs to be done 
at the level of bare lattice operators, and through the mass correction to the renormalization constants. In short, 
\bea
&&A^{\rm impr.}_\mu(x) =Z_A(g_0^2) \bigl( 1 + b_A(g_0^2) (am_q)\bigr) \bigl[ A_\mu(x) + c_A \partial_\mu P(x) \bigr]\,, \nn\\
{\phantom{\huge{l}}}\raisebox{-.01cm}{\phantom{\Huge{j}}}
&& V^{\rm impr.}_\mu(x) =Z_V(g_0^2)\bigl( 1 + b_V(g_0^2) (am_q)\bigr) \bigl[ V_\mu(x) + c_V \partial_\nu T_{\mu \nu}(x) \bigr]\,,
\eea
where $T_{\mu\nu}(x) = i\bar q(x)\sigma_{\mu\nu}q(x)$, and $\sigma_{\mu\nu}=(i/2)[\gamma_\mu,\gamma_\nu]$. 
\begin{table}[t]
\centering
\begin{tabular}{|c|c|c|c|c|c|c|} \hline
{\phantom{\huge{l}}}\raisebox{-.2cm}{\phantom{\Huge{j}}}
{\hspace{-1.5mm}}{ $\beta$}& { $c_{SW}$~\cite{CSW}} & { $c_V^{\rm bpt.}$~\cite{BPT}} & { $Z_V$}~\cite{ZV-QCDSF,ZV-alpha} &  { $Z_A$}~\cite{ZA-QCDSF,ZA-alpha}   &  { $b_V$}~\cite{ZV-QCDSF} &  
{ $b_A^{\rm bpt.}$}~\cite{BPT}  \\ \hline \hline
{\phantom{\Large{l}}}\raisebox{.2cm}{\phantom{\Large{j}}}
{\hspace{-1.5mm}}$5.29$ & $1.919$ & $-0.034$ & $0.743$ & $0.772(4)$ & $1.91$ & $1.31$\\
{\phantom{\Large{l}}}\raisebox{.2cm}{\phantom{\Large{j}}}
{\hspace{-1.5mm}}$5.40$ &  $1.823$ & $-0.032$ & $0.757$ & $0.783(4)$ & $1.79$ & $1.30$ \\
\hline \end{tabular}
\caption{\label{table-RCS}\footnotesize  Renormalization and improvement constants used in this work. Apart from $c_V$ and $b_A$, which are estimated by using the  1-loop (boosted) perturbation theory, all
 the constants are determined non-perturbatively. }
\label{tab:masses}
\end{table}
When sandwiched between the external $V$ and $P$ states, or more specifically $D^\ast$ and $D$, the piece proportional to 
$c_A(g_0^2)$ does not modify the form factors $A_{1,2}(q^2)$. It only changes $A_0(q^2)$ as
\bea
A_0^{\rm impr.}(q^2) = \left( 1 - c_A(g_0^2) {q^2\over 2 m_q^{\rm AWI}} \right)\,,
\eea
where $m_q^{\rm AWI}$ stands for the bare quark mass obtained on the lattice via the axial Ward identity (AWI). In other words, our determination of the coupling $g_{D^\ast D\pi}$ does not 
feel the effect of improvement of the bare axial current. Instead,  the vector form factor does get improved as 
\bea
V^{\rm impr.}(q^2) =V(q^2) + c_V(g_0^2) \bigl( m_D + m_{D^\ast}\bigr) \ T_1(q^2)\,,
\eea
where we used the standard parameterization of the tensor density matrix element, 
\bea
\langle D(k )\vert T^{\mu \nu} \vert D^\ast (p,e_\lambda)\rangle
\!\!&=&\!\!  \left\{ \epsilon^{\mu \nu \alpha \beta} 
\left[
\left(  p_\beta + k_\beta - {m_{D^\ast}^2 - m_{D}^2 \over q^2 } q_\beta \right) T_1(q^2)
+ {m_{D^\ast}^2 - m_{D}^2 \over q^2 } q_\beta 
 T_2(q^2) \right]  \right. \cr
&&\hspace*{-9mm} \left. + {2 k^\alpha \over q^2}  \epsilon^{\mu \nu \sigma \rho } 
k_\sigma p^{\prime}_{\rho} \left( T_2(q^2)- T_1(q^2)
 + {q^2\over m_{D^\ast}^2-m_{D}^2}T_3(q^2) \right) \right\} e^{ \lambda}_\alpha(p)\,.
\eea
We implemented this improvement in our calculation and the results presented in the next section are improved and renormalized. The values of the improvement  ($c_V(g_0^2)$) and 
renormalization constants ($Z_{V,A}(g_0^2)$,  $b_{V,A}(g_0^2)$) used in this work, with the appropriate list of references, are listed in table~\ref{table-RCS}.

\section{Numerical results}

\subsection{Lattice details}
The correlation functions needed to complete this work are computed on the gauge field configurations produced by the QCDSF collaboration~\cite{qcdsf-alg}. They were obtained by using the Wilson plaquette 
gauge field action, and the  improved  Wilson fermions with $N_{\rm f}=2$ mass-degenerate dynamical light quarks at $\beta=5.29$ ($a\approx 0.075$~fm) and 
$\beta=5.40$ ($a\approx 0.065$~fm) on the lattices of volume $24^3\times 48$. From the ensemble of available configurations we chose the ones corresponding 
to the lighter sea quarks, the hopping parameters of which are listed in table~\ref{table-details}. The number of configurations used in this study is also given in the same table. 
\begin{table}[h]
\centering
\begin{tabular}{|c|c|c|c|c|c|} \hline
{\phantom{\huge{l}}}\raisebox{-.2cm}{\phantom{\Huge{j}}}
{\hspace{-2.5mm}}{ $\beta\quad$}&{$\kappa_{\rm sea}=\kappa_{\rm val.}$}&\# conf. & { $m_\pi$ } & { $m_D$} &   $m_{D^\ast}$   \\ \hline \hline
{\phantom{\Large{l}}}\raisebox{.2cm}{\phantom{\Large{j}}}
{\hspace{-2.5mm}}$5.29$ & $\kappa_1=0.1355$ & $60$ & $0.3271(23)$ & $0.789(3)$ & $0.854(4)$  \\
{\phantom{\Large{l}}}\raisebox{.2cm}{\phantom{\Large{j}}}
{\hspace{-2.5mm}}             & $\kappa_2=0.1359$ & $80$ & $0.2451(25)$ & $0.760(5)$ & $0.805(6)$  \\
{\phantom{\Large{l}}}\raisebox{.2cm}{\phantom{\Large{j}}}
{\hspace{-2.5mm}}  & $\kappa_3=0.1362$ & $100$ & $0.1549(24)$ & $0.731(6)$ & $0.783(9)$  \\ \hline
{\phantom{\Large{l}}}\raisebox{.2cm}{\phantom{\Large{j}}}
{\hspace{-2.5mm}}$5.40$ & $\kappa_1=0.1356$& $130$  & $0.3124(18)$ & $0.725(3)$ & $0.771(3)$  \\
{\phantom{\Large{l}}}\raisebox{.2cm}{\phantom{\Large{j}}}
{\hspace{-2.5mm}}  & $\kappa_2=0.1361$ & $120$ & $0.2166(23)$ & $0.691(4)$ & $0.734(5)$  \\
{\phantom{\Large{l}}}\raisebox{.2cm}{\phantom{\Large{j}}}
{\hspace{-2.5mm}}  & $\kappa_3=0.13625$ & $160$ & $0.1843(26)$ & $0.681(2)$ & $0.721(4)$  \\ \hline
 \end{tabular}
\caption{\label{table-details}\footnotesize  \sl Masses of the light-light and heavy-light mesons in which the light valence quark has the same mass as the sea quark.  All results are given in lattice units and the charm quark hopping parameter is fixed to $\kappa_{\rm charm}^{[\beta =5.25]} = 0.125$ and $\kappa_{\rm charm}^{[\beta =5.40]} = 0.126$. The gauge field configurations are separated by $20$ unit-length HMC trajectories. }
\end{table}
We chose to use the configurations that are separated by $20$ molecular dynamics trajectories of the unit length in the hybrid Monte-Carlo algorithm (HMC). In computing the quark propagators we attempted 
using the so called ``color dilution" technique~\cite{color-dilution} but since we did not observe any appreciable effect in improving the statistical quality of our correlation functions, we returned to the standard BiCG-Stab 
algorithm~\cite{bicg}. With the computed propagators we then calculated the $2$- and $3$-point correlation functions in a standard way. We first recomputed the masses of light quarks and of the pseudoscalar 
mesons consisting of the light valence quark and antiquark the mass of which is  equal to that of the sea quark. 

We tried to fix the charm quark mass in several ways, of which we describe  one. 
Since the charm quark is not propagating in the sea, its phenomenology is quenched and its hopping parameter can be fixed by tuning its value to make the ratio of the light-strange pseudoscalar and the strange-charmed 
mesons equal to its physical value,  $m_{\eta_{ss}}/m_{D_s}=0.35$, where $m_{\eta_{ss}} = \sqrt{2 m_K^2 - m_\pi^2}=0.684$~GeV. 
On the very same configurations the strange quark mass has been computed in ref.~\cite{strange}, quoting $r_0 m_s^{\overline{\rm MS}}(2~\gev ) = 0.2545(47)$ at $\beta=5.29$, and  
$r_0 m_s^{\overline{\rm MS}}(2~\gev ) = 0.2671(48)$ at $\beta = 5.40$. By using these values and the numerical constants provided in ref.~\cite{strange}, as well as the force parameter~\cite{R0}, 
$r_0/a = 6.25(10)_{\beta=5.29}$, $7.39(26)_{\beta=5.40}$, we extract $\kappa_{\rm strange}^{[\beta=5.29]}\approx 0.1355$ and 
$\kappa_{\rm strange}^{[\beta =5.40]}\approx 0.1359$. After having fixed one valence quark to $\kappa_{\rm strange}$ and  varying $\kappa_{\rm charm}$, we were able to  reproduce $m_{\eta_{ss}}/m_{D_s}=0.35$, 
 for $\kappa_{\rm charm}^{[\beta =5.25]} = 0.125$, and $\kappa_{\rm charm}^{[\beta =5.40]} = 0.126$, after linearly extrapolating the sea quark mass to zero. By adopting other strategies to fix the charm hopping 
 parameter, the resulting  $\kappa_{\rm charm}$ differs by at most $5\times 10^{-4}$. We checked, however, that the results of this paper are insensitive to that variation. In table~\ref{table-details} we list the pseudoscalar 
($D$)  and vector ($D^\ast$)  meson masses, for which the light quark hopping parameter $\kappa_q=\kappa_{\rm sea}$ and the heavy quark is fixed to $\kappa_{\rm charm}$. Notice the labels $\kappa_{1,2,3}$ in table~\ref{table-details}, which will be used in the next subsections when reporting our results for the form factors. Every time the light valence quark appears in a correlation function, its mass is always kept equal to the sea quark mass.

\subsection{Form factors}

\begin{figure}[t!]
\begin{center}
\hspace*{-8mm}\epsfig{file=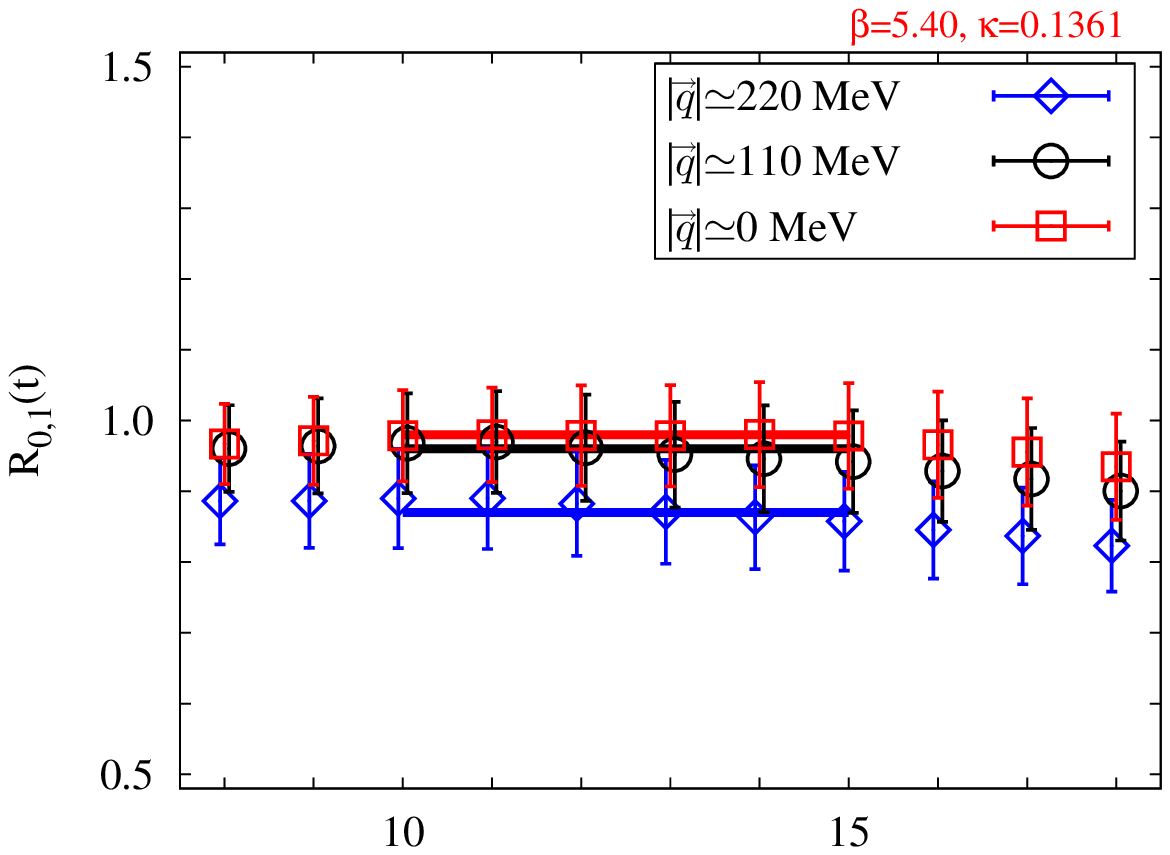, height=6.3cm}~\epsfig{file=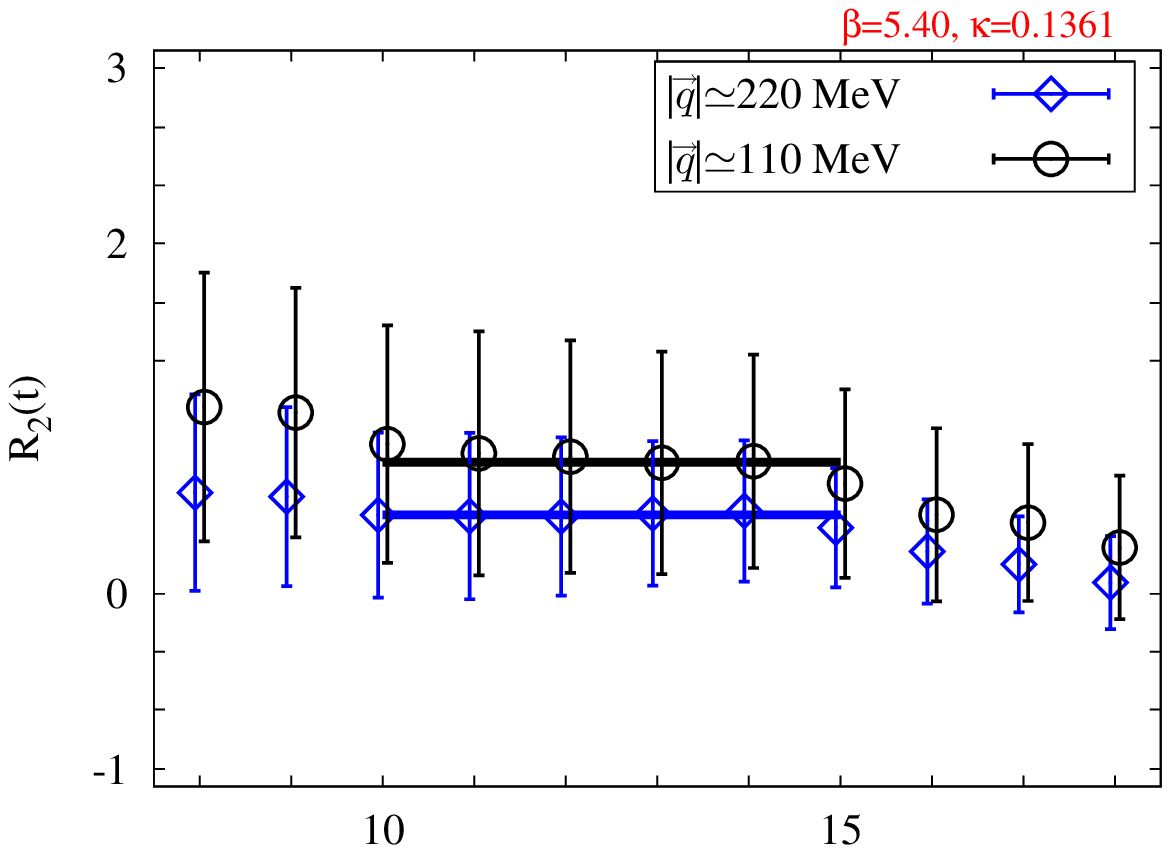, height=6.1cm}\\ 
\vspace*{5mm}\hfill \\
\hspace*{-7mm}\epsfig{file=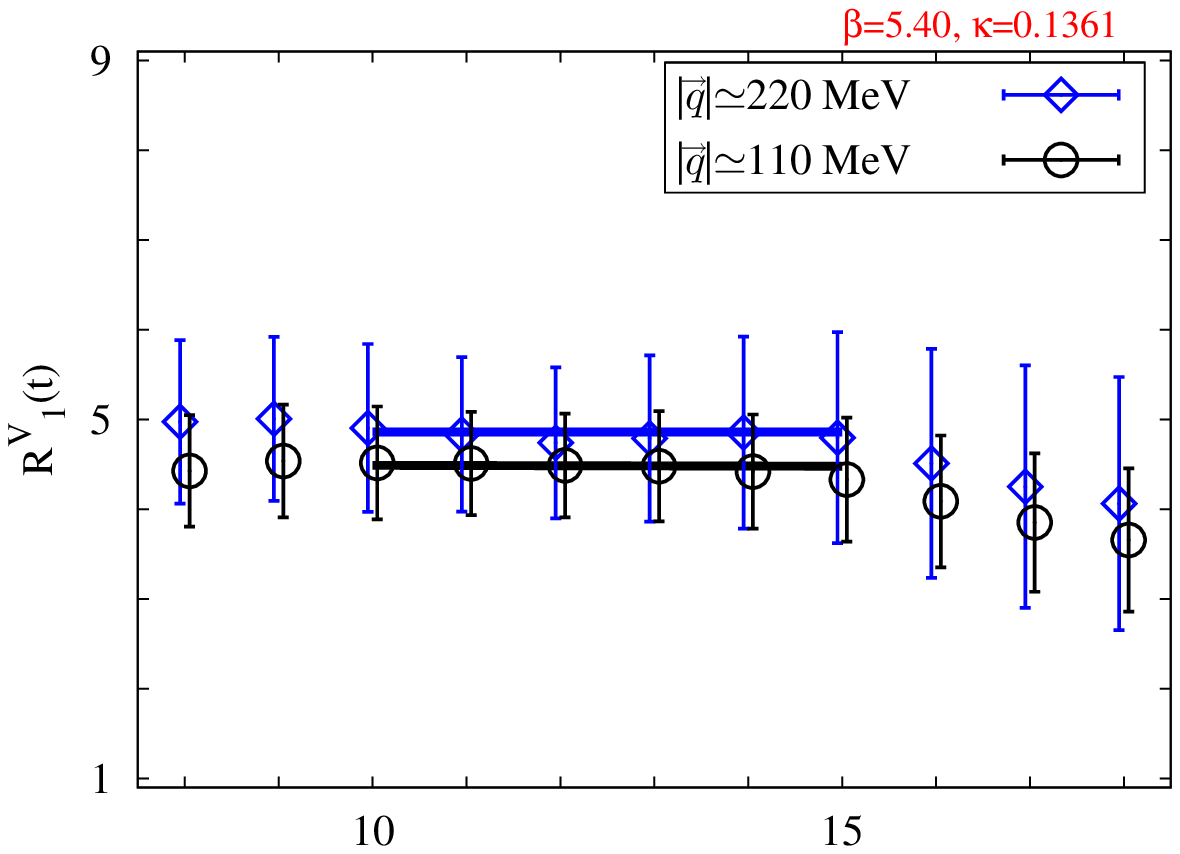, height=6.2cm}~\epsfig{file=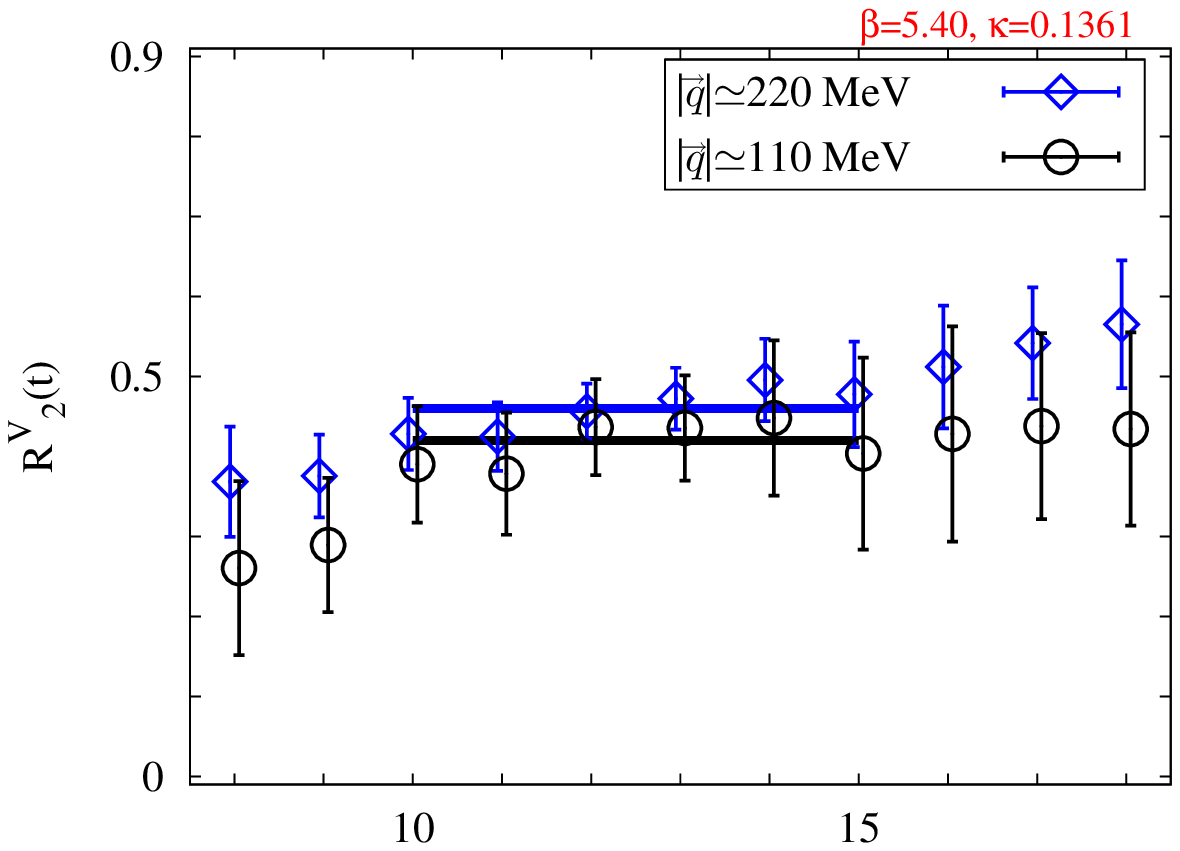, height=6.2cm}
\caption{\label{figR12}\footnotesize{\sl 
In the upper plots we illustrate the time dependence of the ratios $R_0(t)$, $R_1(t)$ and $R_2(t)$, defined in eqs.(\ref{defR0},\ref{defR1},\ref{defR2}), leading to the dominant form factor $A_1(q^2)$ and 
to  $A_2(q^2)/A_1(q^2)$. In the lower plots the similar illustration is provided for the ratios $R^V_1(t)$ and $R^V_2(t)$, defined in eqs.(\ref{RV1}) and (\ref{RV2}), and leading to  the form factor $V^{qq}(q^2)$ and $V^{cc}(q^2)/V^{qq}(q^2)$, respectively. The fit on the plateau interval $t\in[10,15]$ is also shown, for both $\theta_0=0.5$ and $\theta_0=1.0$, corresponding to about $110$~MeV and $220$~MeV respectively.}} 
\end{center}
\end{figure}
In this subsection we present our main numerical results. We first computed the ratios (\ref{defR0}) which exhibit the long plateaus and lead to $A_1(q^2_{\rm max})$, with $q^2_{\rm max}  = (m_{D^\ast}-m_D)^2$. 
Since $q^2_{\rm max}$ is very small and since the shape of the axial form factor is expected to be driven by a rather heavy $a_1$-state,  the result for $A_1(0)$ will differ only slightly with respect to $A_1(q^2_{\rm max})$.~\footnote{
 From the light-light two point correlation functions, $C_{AiAi}(t)=\sum_{\vec x}\langle A_i(0) A_i(x)$ we extract the masses of $a_1$-state and from our data we have:
\bea
(m_{a_1}/m_\pi) = \{1.9(2),2.5(3),3.9(3)\}_{\beta=5.29},\, {\rm and}\quad (m_{a_1}/m_\pi) = \{2.0(0),2.6(1),2.9(1)\}_{\beta=5.40}. \nn
\eea
On our lattices, $m_{a_1}$ is about $10$ times larger than $q^2_{\rm max}$.}
To control the interpolation to $q^2=0$ we evaluate the ratios (\ref{defR2}) by chosing the twisting angle $\theta_0 \in \{ 0.5,1.0,1.5\}$, which are sufficient to make $q^2$ straddle around zero for all of our $\kappa_{1,2,3}$. 

From our data we observe that the quality of the signal for the ratio $R_{1}(t)$ remains good when $\theta_0\neq 0$ and small, whereas the signal for $R_{2}(t)$ is bad. The illustration of the ratios $R_{1,2}(t)$ 
is provided in fig.~\ref{figR12}.  
Whenever the contribution of the form factor $A_2(q^2)$ to a correlation function is significant,  those correlation functions are numerically much smaller than those in which $A_2$ does not contribute, e.g. 
$ \vert \widetilde C^{(2)}_{0i}(\vec q;t)\vert \ll  \vert \widetilde C^{(2)}_{ii}(\vec q;t)\vert $. 
\begin{table}[t!]
\begin{center} 
\hspace*{-.9cm}{\scalebox{.9}{\begin{tabular}{c|c|c|c|c|c|c|c|}
\cline{3-8}
\multicolumn{2}{l|}{{\phantom{\Huge{l}}}\raisebox{-.1cm}{\phantom{\Huge{j}}}}& \multicolumn{3}{c|}{${\beta=5.29}$} &   \multicolumn{3}{c|}{$\beta=5.40$} \\
\cline{3-8}
\multicolumn{2}{l|}{}&{\phantom{\Huge{l}}}\raisebox{-.1cm}{\phantom{\Huge{j}}} $(r_0 q)^2$ & $\quad A_1(q^2)\quad $ & $g_c(q^2,m_\pi^2)$ &$(r_0 q)^2$ & $\quad A_1(q^2)\quad $& $g_c(q^2,m_\pi^2)$  \\
 \cline{2-8} 
{\phantom{\Huge{l}}}\raisebox{-.1cm}{\phantom{\Huge{j}}}
{$\theta_0=0.0$}  & $\kappa_1$ &$0.17(2)$  & $1.40(11)$ & $1.45(11)$ &  $ 0.12(1)$ &  $1.11(4)$ &  $1.15(4)$\\  
{\phantom{\Huge{l}}}\raisebox{-.1cm}{\phantom{\Huge{j}}}
            & $\quad\kappa_2\quad$ &$0.09(3)$ & $1.06(10)$ & $1.09(11)$ &  $0.11(2)$&   $0.98(7)$ &  $1.01(7)$ \\ 
{\phantom{\Huge{l}}}\raisebox{-.1cm}{\phantom{\Huge{j}}}
                                & $\kappa_3$ & $0.10(3)$ & $0.93(14)$ & $0.95(15)$ & $0.09(1)$ &  $0.90(3)$ &  $0.93(3)$\\ \cline{2-8} 
{\phantom{\Huge{l}}}\raisebox{-.1cm}{\phantom{\Huge{j}}}

{$\theta_0=0.5$}  & $\kappa_1$ &$0.11(2)$  & $1.37(11)$ & $1.42(19)$ &  $ 0.04(1)$ &  $1.09(4)$ &  $1.10(7)$\\  
{\phantom{\Huge{l}}}\raisebox{-.1cm}{\phantom{\Huge{j}}}
            & $\quad\kappa_2\quad$ &$0.03(3)$ & $1.04(10)$ & $0.98(17)$ &  $0.03(2)$&   $0.96(7)$ &  $1.04(2)$ \\ 
{\phantom{\Huge{l}}}\raisebox{-.1cm}{\phantom{\Huge{j}}}
                                 & $\kappa_3$ & $0.04(3)$ & $0.89(13)$ & $0.88(26)$ & $0.02(1)$ &  $0.88(3)$ &  $0.91(13)$\\ \cline{2-8} 
{\phantom{\Huge{l}}}\raisebox{-.1cm}{\phantom{\Huge{j}}}

{$\theta_0=1.0$} & $\kappa_1$   & $-0.05(2)$ & $1.29(10)$ & $1.34(11)$ & $-0.19(1)$&  $1.03(4)$ &  $1.05(4)$ \\  
{\phantom{\Huge{l}}}\raisebox{-.1cm}{\phantom{\Huge{j}}}
                                & $\kappa_2$ & $-0.13(2)$ & $0.96(10)$ & $0.99(13)$ & $-0.20(2)$ &  $0.87(7)$ &  $0.92(8)$\\ 
{\phantom{\Huge{l}}}\raisebox{-.1cm}{\phantom{\Huge{j}}}
                               & $\kappa_3$ & $-0.12(3)$ & $0.85(18)$& $0.83(17)$ & $-0.21(1)$&   $0.79(3)$ &  $0.80(3)$ \\ \cline{2-8} 
{\phantom{\Huge{l}}}\raisebox{-.1cm}{\phantom{\Huge{j}}}

{$\theta_0=1.5$} & $\kappa_1$   & $-0.33(2)$ & $1.16(10)$ & $1.21(10)$ &  $-0.56(1)$ &  $-$ & $-$  \\  
{\phantom{\Huge{l}}}\raisebox{-.1cm}{\phantom{\Huge{j}}}
                                 & $\kappa_2$ & $-0.40(2)$ & $0.85(9)$ & $0.88(11)$ & $-0.57(1)$&  $-$ &  $-$  \\ 
{\phantom{\Huge{l}}}\raisebox{-.1cm}{\phantom{\Huge{j}}}
                                 & $\kappa_3$ & $-0.39(3)$ &$0.66(14)$ & $0.64(12)$ & $-0.58(1)$ &  $-$ &$-$  \\  \cline{2-8}
\multicolumn{0}{l}{}
{\phantom{\Huge{l}}}\raisebox{-.1cm}{\phantom{\Huge{j}}}
\\
\cline{3-8}
\multicolumn{2}{l|}{}&{\phantom{\Huge{l}}}\raisebox{-.1cm}{\phantom{\Huge{j}}} $(r_0 q)^2$ & $V^{qq}(q^2)$& $\quad\displaystyle{\frac{V^{cc}(q^2)}{V^{qq}(q^2)}}\quad$ &$(r_0 q)^2$ & $V^{qq}(q^2)$& $\quad\displaystyle{\frac{V^{cc}(q^2)}{V^{qq}(q^2)}}\quad$  \\
 \cline{2-8} 
{\phantom{\Huge{l}}}\raisebox{-.1cm}{\phantom{\Huge{j}}}
{$\theta_0=0.5$}  & $\kappa_1$ &$0.11(2)$  & $3.73\pm 1.63$ & $0.03(39)$ &  $ 0.04(1)$ &  $5.03\pm 0.59$ &  $0.50(9)$\\  
{\phantom{\Huge{l}}}\raisebox{-.1cm}{\phantom{\Huge{j}}}
            & $\quad\kappa_2\quad$ &$0.03(3)$ & $2.80\pm 1.39$ & $0.89(48)$ &  $0.03(2)$&   $4.83\pm 0.88$ &  $0.42(9)$ \\ 
{\phantom{\Huge{l}}}\raisebox{-.1cm}{\phantom{\Huge{j}}}
            & $\kappa_3$ & $0.04(3)$ & $5.69\pm 3.34$ & $0.32(45)$ & $0.02(1)$ &  $4.35\pm 1.12$ &  $0.42(17)$\\ \cline{2-8} 
{\phantom{\Huge{l}}}\raisebox{-.1cm}{\phantom{\Huge{j}}}

{$\theta_0=1.0$} & $\kappa_1$   & $-0.05(2)$ & $4.82\pm 0.89$ & $0.34(8)$ & $-0.19(1)$&  $4.73\pm 0.34$ &  $0.50(5)$ \\  
{\phantom{\Huge{l}}}\raisebox{-.1cm}{\phantom{\Huge{j}}}
            & $\kappa_2$ & $-0.13(2)$ & $4.38\pm 0.74$ & $0.51(10)$ & $-0.20(2)$ &   $4.48\pm 0.59$ &  $0.46(5)$\\ 
{\phantom{\Huge{l}}}\raisebox{-.1cm}{\phantom{\Huge{j}}}
            & $\kappa_3$ & $-0.12(3)$ & $6.07\pm 2.02$ & $0.35(20)$ & $-0.21(1)$&   $4.15\pm 0.64$ &  $0.42(8)$ \\ \cline{2-8} 
{\phantom{\Huge{l}}}\raisebox{-.1cm}{\phantom{\Huge{j}}}
{$\theta_0=1.5$} & $\kappa_1$   & $-0.33(2)$ & $4.76\pm 0.68$ & $0.41(6)$ &  $-0.56(1)$ &  $-$ & $-$  \\  
{\phantom{\Huge{l}}}\raisebox{-.1cm}{\phantom{\Huge{j}}}
            & $\kappa_2$ & $-0.40(2)$ & $4.45\pm 0.55$ & $0.47(7)$ & $-0.57(1)$&  $-$ &  $-$  \\ 
{\phantom{\Huge{l}}}\raisebox{-.1cm}{\phantom{\Huge{j}}}
            & $\kappa_3$ & $-0.39(3)$ &$4.70\pm 1.00$ & $0.38(10)$ & $-0.58(1)$ &  $-$ &$-$  \\ \cline{2-8} 
\end{tabular} }}
\caption{\label{tableA1}
\footnotesize The form factors  $A_1(q^2)$ and $g_c(q^2,m_\pi^2)$ [c.f. eq.(\ref{gcc})] relevant to the axial coupling, and the form factors $V^{qq}(q^2)$ and   $V^{cc}(q^2)/V^{qq}(q^2)$ relevant to the magnetic moment of our $D^\ast$-meson.  }
\end{center}
\vspace*{-.3cm}
\end{table}
As a result the extracted information on  $A_2(q^2)$ suffers from  large statistical errors. We attempted several other 
options to extract $A_2(q^2)/A_1(q^2)$ but none appeared to be better than the one based on using eq.~(\ref{defR2}), the results of which we present here.  Luckily, however, the ratio $A_2(q^2)/A_1(q^2)$ in eq.~(\ref{defA}) comes only as a small correction to the dominant $A_1(q^2)$-contribution, i.e. it is suppressed by $(m_{D^\ast}-m_D)/(m_{D^\ast}+m_D)$. 
 In table~\ref{tableA1} we list our results for the form factor $A_1(q^2)$ as obtained from the fit to $R_{0,1}(t)$ to a constant between $t\in [10,15]$. The results for $A_2(q^2)/A_1(q^2)$ are combined into the quantity 
\bea\label{gcc}
g_c(q^2,m_\pi^2) ={ 2\sqrt{m_D m_{D^\ast}} \over m_D + m_{D^\ast} }A_1(q^2)\left[1+{m_{D^\ast}-m_D\over m_{D^\ast}+m_D}{A_2(q^2)\over A_1(q^2)}\right]  \,,
\eea
that will eventually lead us to  the $g_c$-coupling,  defined as,
\bea
g_{D^\ast D\pi} ={ 2\sqrt{m_D m_{D^\ast}}\over f_\pi} g_c\;.
\eea
Concerning the  momentum injections corresponding to $\theta_0\in \{0.5,1.0,1.5\}$, by chosing $r_0=0.467$~fm, we have $\vert\,\!\vec q\,\vert_{\beta =5.29}\in \{ 96(2),191(3),287(5)\}$~MeV, and 
$\vert\,\!\vec q\,\vert_{\beta =5.40}\in \{ 113(4)$, $226(8)$, $339(12)\}$~MeV. These values would be $7\%$ smaller if we made another (also standard) a choice, namely $r_0=0.5$~fm. 
The results for all our data-points (both $\beta$, all $\kappa_{\rm sea}$, and all $\theta_0$) are collected in table~\ref{tableA1}. Notice that at $\beta=5.40$ we did not compute the form factors 
with $\theta_0=1.5$, as it was clear that for that value  $q^2 \ll 0$ and would not help us interpolating to $q^2=0$.

We now turn to the results for the vector form factor. In this case the form factor is not accessible when both mesons are at rest. When $\theta_0\neq 0$ we obtain the results for $V^{qq}(q^2)$ 
and $V^{cc}(q^2)/V^{qq}(q^2)$ from the fit of the ratios $R^V_1(t)$ and $R^V_2(t)$ to a constant, on the plateau which we chose to be $t_{\rm fit}\in [10,15]$, as before. The quality of the signal is 
shown in fig.~\ref{figR12}, and the complete list of results is provided in table~\ref{tableA1}. 
From that table we see that, contrary to what one would naively expect,  the charmed contribution to the electromagnetic current is not negligible when compared 
to the light quark piece. In other words around $30\%$ of the magnetic moment of the $D^{\ast 0}$-meson comes from charm. That is an indication that indeed for some quantities, such as the case at hand,  
the leading term in heavy quark expansion is a  bad approximation of the charmed hadrons properties.~\footnote{The contribution to the magnetic moment of the heavy-light vector meson that comes from  the infinitely heavy quark is  zero.}

\subsection{Towards the physical results}

\begin{figure}
\begin{center}
\epsfig{file=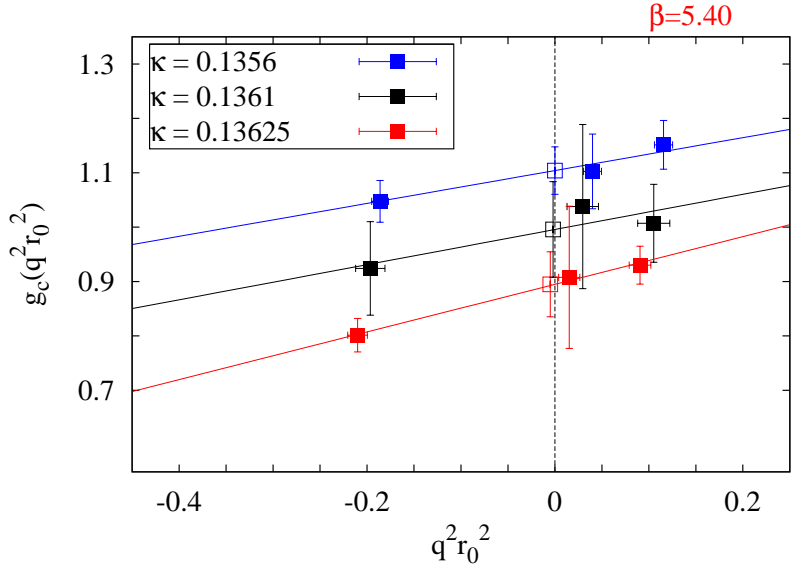, height=7.8cm}
\caption{\label{fig:extrap}\footnotesize{\sl 
Linear interpolation in $q^2$ to $q^2=0$. Illustration is provided with the unquenched data obtained at $\beta=5.40$ with $N_{\rm f}=2$, which are denoted by the full symbols. The empty symbols show the results of interpolation for each of our three light quark (corresponding pion) masses. We stress that the results are obtained with the fully unquenched light quark.}} 
\end{center}
\end{figure}

To reach the physical results, we need to make two important steps. First we need to interpolate to $q^2=0$, which we can do either linearly (or quadratically), or by using some physical assumption, 
such as the nearest pole dominance of the particle exchanged in the $t$-channel, namely $a_1$-meson  to $g_c(q^2,m_\pi^2)$, and the 
$\rho$-meson dominance  to the $V^{qq}(q^2)$. The  masses of $a_1$ and $\rho$ are easily computed on the same lattice. However, since the range of $q^2$'s that we are probing is very short and 
since our results are not sufficiently accurate we cannot distinguish amongst different interpolating formulas,  and we choose to quote our value obtained through the linear interpolation to $q^2=0$. 
All results are listed, in table~\ref{q2=0}, and the illustration is provided in fig.~\ref{fig:extrap} 
for the case of $g_c(q^2,m_\pi^2)$. 
\begin{table}[t] 
\centering
\hspace*{-.7cm}{\scalebox{.88}{\begin{tabular}{|c|ccc|ccc|}  
\cline{2-7}
\multicolumn{1}{l|}{ {\phantom{\huge{l}}}\raisebox{-.2cm}{\phantom{\huge{j}}} }
 &  \multicolumn{3}{c|}{$\beta=5.29$}  &   
\multicolumn{3}{c|}{$\beta=5.40$} \\ 
\cline{2-7}
\multicolumn{1}{l|}{ {\phantom{\huge{l}}}\raisebox{-.2cm}{\phantom{\huge{j}}} }
&{ $\kappa_1$}   & {$\kappa_2$} & {$\kappa_3$} &  { $\kappa_1$}   & {$\kappa_2$} & {$\kappa_3$}  \\ \hline  \hline 
{\phantom{\huge{l}}}\raisebox{-.2cm}{\phantom{\huge{j}}}
\hspace*{-5mm}$V^{qq}(0)$ & $4.56\pm 1.12$ & $3.14\pm 1.43$ & $5.88\pm 2.88$ & $4.98\pm 0.54$&  $4.78\pm 0.82$ & $4.34\pm 1.08$  \\ 
{\phantom{\huge{l}}}\raisebox{-.2cm}{\phantom{\huge{j}}}
\hspace*{-5mm}$V^{cc}(0)/V^{qq}(0)$ &  $0.26\pm 0.13$ &  $0.76\pm 0.36$ &   $0.33\pm 0.36$ &  $0.50\pm 0.08$&   $0.43\pm 0.08$ &  $0.42\pm 0.16$ \\  \hline 
{\phantom{\huge{l}}}\raisebox{-.2cm}{\phantom{\huge{j}}}
\hspace*{-5mm}$F_u(0,m_\pi^2)$ & $3.62\pm 1.17$ & $3.71\pm 1.57$ & $5.19\pm 2.14$ & $4.98\pm 0.45$&  $4.55\pm 0.81$ & $4.10\pm 0.89$ \\ 
{\phantom{\huge{l}}}\raisebox{-.2cm}{\phantom{\huge{j}}}
\hspace*{-5mm}$F_d(0,m_\pi^2)$ & $-0.95\pm 0.63$ &  $-0.56\pm 0.77$ &   $-0.69\pm 1.55$ &  $-0.00\pm 0.26$&   $-0.23\pm 0.27$ &  $-0.24\pm 0.52$ 
\\ \hline  \hline 
{\phantom{\huge{l}}}\raisebox{-.2cm}{\phantom{\huge{j}}}
\hspace*{-5mm}$g_c(0,m_\pi^2)$ & $1.37\pm 0.11$ &  $1.02\pm 0.12$ &   $0.89\pm 0.14$ &  $1.10\pm 0.04$&   $0.99\pm 0.07$ &  $0.90\pm 0.06$ 
\\ \hline 
\end{tabular} }}
\caption{\label{q2=0} {\footnotesize \sl Results of the linear interpolation of our results from table~\ref{tableA1} to $q^2=0$, for each of the light quarks that we have at our disposal ($\kappa_{1,2,3}$). We also show the particular combinations $F_{u,d}(0,m_\pi^2)$ which are needed to arrive at the magnetic moments, i.e. to the couplings  $g_{D^{\ast +} D^+ \gamma}$ and $g_{D^{\ast 0}  D^0 \gamma}$.
}}
\label{tab:fds}
\end{table}
In that same table  we also give the results for  $F_{u,d}(0,m_\pi^2)$ obtained after combining the interpolated values of the vector form factors as indicated in eq.~(\ref{defV}). 

Second step is far more delicate. In order to reach the physically interesting result from our calculation an extrapolation to the chiral limit is required. 
The relevant formula, derived in HMChPT, reads~\cite{jernej,stewart} 
\bea\label{chpt-g}
g_c(m_\pi^2) = g_c^{(0)}\left[  1 - {4 (g_c^{(0)})^2 \over (4 \pi f r_0)^2 } (m_\pi r_0)^2 \log(m_\pi r_0)^2 + {c_g\over r_0^2} (m_\pi r_0)^2 \right]\,,
\eea
with the particularly interesting parameter $g_c^{(0)}$ which appears in all the HMChPT formulas that are being used to guide the chiral extrapolations 
of the phenomenologically interesting quantities in charm physics computed on the lattice. In the expression~(\ref{chpt-g}) we multiplied with suitable factors of $r_0/a$ to make the expression independent 
on the lattice spacing, $a$. 
Two important underlying assumptions are made when applying this formula: (i) Eq.~(\ref{chpt-g}) is obtained in the static heavy quark 
limit and its use is equivalent to the assumption that the $1/m_c^{n\geq 1}$-corrections do not modify the chiral behavior of this coupling. That assumption is quite reasonable because the chiral 
dynamics is driven by the light quark vertex (in our case the current $\bar q\gamma_\mu \gamma_5 q$), while the heavy quark is only a spectator; (ii) The HMChPT formula applies to our data too. Of our three light quarks in each set, one is heavier than the physical strange 
quark, one is somewhat lighter and our lightest dynamical quark is about a third (half) of the mass of the strange quark when $\beta=5.29$ ($5.40$). The assumption that such a heavy light quarks obey the behavior
driven by the HMChPT formula is strong and the check of its validity is one of the most active research topics in the lattice QCD community. 
With these two underlying assumption in mind we fit our data to eq.~(\ref{chpt-g})  and obtain 
\bea
g^{(0)}_{\rm HMChPT} = \{ 0.63(11)_{\beta=5.29}, 0.68(7)_{\beta=5.40}\}\,, 
\eea
while  at the physical pion mass we have, $g_c(m^{{\rm phys}\ 2}_\pi) =  \{ 0.66(11)_{\beta=5.29}, 0.71(7)_{\beta=5.40}\} $. 
If the logarithmic correction in eq.~(\ref{chpt-g})  is set to zero we obtain the simple linear extrapolation which gives 
$g^{(0)}_{\rm linear} = \{ 0.70(15)_{\beta=5.29}, 0.81(11)_{\beta=5.40}\}$.  As a small check, we combine the data of two sets and use only those obtained in the case in which the sea quark mass is lighter than the physical 
strange quark mass, and we obtain $g^{(0)}_{\rm HMChPT} = 0.67(14)$, and $g_c(m^{{\rm phys}\ 2}_\pi) = 0.69(14)$. 
These results are to be compared with the quenched value obtained in ref.~\cite{Abada:2002xe}, $g_c =0.66(8)(5)$. We conclude that no effect due to unquenching is visible from our results.

On the other hand, the result for the coupling to the soft photon is completely new. It has never been computed even in the quenched approximation. 
HMChPT formula for this coupling was obtained in refs.~\cite{stewart,cheng} and reads
\bea\label{cheng}
{\phantom{\huge{l}}}\raisebox{-.6cm}{\phantom{\huge{j}}}
&&F_u(m_\pi^2) =  \sqrt{m_D m_{D^\ast}} \left[ {2\over 3}{ \beta}\left(1 - {3\over 2} {3+ g_c^2 \over (4 \pi f r_0)^2 } (m_\pi r_0)^2 \log(m_\pi r_0)^2\right) - {g_c^2 + d_u\over 4 \pi f^2} m_\pi  \right]\,,\cr
&&F_d(m_\pi^2) = - \sqrt{m_D m_{D^\ast}} \left[ {1\over 3} \beta \left(1 - {3\over 2} {1+ 2 g_c^2 \over (4 \pi f r_0)^2 } (m_\pi r_0)^2 \log(m_\pi r_0)^2\right) - {g_c^2+d_d \over 4 \pi f^2} m_\pi   \right],
\eea
with $\beta$-being the dimension-full parameter which is a measure of the magnetic moment of the light quark inside the static heavy-light vector meson, and $d_{u,d}$ are the counter-term coefficients which are 
supposed to be obtained from the fit.  As it can be seen from table~\ref{q2=0}, our data 
for $F_d(0,m_\pi^2)$ are consistent with zero, whereas the results for $F_u(0,m_\pi^2)$ change very little when varying the pion mass, contrary to what  the  HMChPT would suggest. 
In other words, the part linear in $m_\pi$ in eq.~(\ref{cheng}) is far too large to provide a suitable description of our data, unless one accepts a huge value for the counter-term coefficient  ($d_{u,d}$) which would then 
contradict  the perturbative feature of HMChPT.  For that reason we only linearly extrapolate our data and arrive at
\bea
{\phantom{\huge{l}}}\raisebox{-.4cm}{\phantom{\huge{j}}}
&&F_u(0) \in \{ (5.3\pm 2.3)_{\beta=5.29}, (3.9\pm 1.1)_{\beta=5.40}\}\,,\cr
&&F_d(0) \in \{ -(0.1\pm 1.4)_{\beta=5.29}, -(0.4\pm 0.6)_{\beta=5.40}\}\,.
\eea
We hope to revisit this issue when more accurate values for $F_u(m_\pi^2)$ obtained with ever lighter $m_\pi$ become available.
\section{Phenomenology}
In this section we discuss our physical results and make a phenomenological discussion based on the currently available experimental information. 
Furthermore, we compare our results for the magnetic moment with the results obtained by other theoretical approaches such as QCD sum rules and quark models.
\subsection{Our results}
From our results obtained with HMChPT we conclude
\bea
g_c \in \{ 0.66(11)_{\beta=5.29}, 0.71(7)_{\beta=5.40}\}\, \Rightarrow  g_{D^\ast D\pi^+}  \in \{ (18.6\pm 3.2)_{\beta=5.29}, (20.1\pm 2.1)_{\beta=5.40}\}\ .
\eea
The expression for the decay width reads
\bea
\Gamma(D^{\ast}\to D\pi) = {C \over 24 \pi m_{D^{\ast }}^2}\  g_{D^\ast D\pi}^2 \vert \vec k_\pi\vert^3\,,
\eea
where $C=1$ if the outgoing pion is charged, and $C=1/2$ if it is neutral. For the charged decaying meson we have~\footnote{
For convenience, in addition to $f_\pi = 131$~MeV, we  list the numerical values of the meson masses used in this paper~\cite{PDG}:
\bea
&& m_{D^+}=1869.6(2)~\mev\,,\quad m_{D^{\ast +}}=2010.3(2)~\mev\,,\quad m_{\pi^+}=139.6~\mev\,,\cr
&& m_{D^0}=1864.8(2)~\mev\,,\quad m_{D^{\ast 0}}=2007.0(2)~\mev\,,\quad m_{\pi^0}=135.0~\mev\,.\nn
\eea  }  
 
\bea
 \vert \vec k_\pi\vert =\sqrt{\left({m_{D^\ast}^2-m_D^2+m_\pi^2\over 2 m_{D^\ast} }\right)^2 - m_\pi^2}~ \Rightarrow \,  \vert \vec k_{\pi^-}\vert = 39.4\ \mev,\,  \vert \vec k_{\pi^0}\vert = 38.3\ \mev ,
\eea
and we get
\bea
\Gamma(D^{\ast -}\to D \pi) =   
\left\{
\begin{array}{ccc} 
(72\pm 24)_{\rm \pi^-}\ \kev ,\quad  (33\pm 11)_{\rm \pi^0}\ \kev \,, &   &  \beta=5.29\ , \\
&& \\
(82\pm 17)_{\rm \pi^-}\ \kev ,\quad  (38\pm 8)_{\rm \pi^0}\ \kev \,, &   &  \beta=5.40\ .
\end{array}
\right.\;
\eea
As for the couplings to the soft photon, our results are
\bea
{\phantom{\huge{l}}}\raisebox{-.32cm}{\phantom{\huge{j}}}
g_{D^{\ast +} D^+\gamma}&=&  \{ -0.1(7)_{\beta=5.29}, -0.2(3)_{\beta=5.40}\}~\gev^{-1}\,,\cr
g_{D^{\ast 0} D^0\gamma} &=& \{ (2.7\pm 1.2)_{\beta=5.29}, (2.0\pm 0.6)_{\beta=5.40}\}~\gev^{-1}\,.
\eea
In other words, the decay width for the radiative decay of a charged $D^\ast$-meson is negligibly small (consistent with zero) whereas for the neutral one we have 
\bea
\label{eq:2}
\Gamma(D^{\ast  0}\to D^0\gamma) =    {\alpha_{\rm em}\over 3} g_{D^{\ast 0} D^0\gamma}^2\, k_{\gamma}^3 =
\left\{
\begin{array}{ccc} 
(48\pm 31)\ \kev \,, &   &  \beta=5.29\ , \\
&& \\
(27\pm 14)\ \kev \,, &   &  \beta=5.40\  ,
\end{array}
\right.\;
\eea
where $2 m_{D^{\ast 0}}\ k_{\gamma }^2= m_{D^{\ast 0}}^2 - m_{D^0}^2$, i.e.  $k_{\gamma }=137.1$~MeV.

\subsection{Comparison with experiment, other methods and more phenomenology} 
\begin{itemize}
\item Let us remind the reader that from the width of the charged $D^\ast$-meson measured by CLEO,  $\Gamma(D^{\ast +})=96\pm 22$~keV~\cite{cleo}, 
and  by using the experimentally established $B(D^{\ast +}\to D^+\gamma) = 0.016(4)$~\cite{PDG}, we get
\bea\label{g-cleo}
{\phantom{\huge{l}}}\raisebox{-.32cm}{\phantom{\huge{j}}}
\hspace*{-12mm}\Gamma^{\rm exp.}(D^{\ast +}) \left[1 - 0.016(4)\right] &=& \Gamma(D^{\ast +} \to D^0 \pi^+) +   \Gamma(D^{\ast +} \to D^+ \pi^0) \hfill \cr
&=& { 2 m_{D^0} |\vec k_{\pi^+}|^3  + m_{D^+}  |\vec k_{\pi^0}|^3\over 12 \pi m_{D^{\ast +}} f_\pi^2 }\  g_c^2\,\Rightarrow \ g_c=0.61(7)\,,
\eea
a well known result which is in good agreement with the value obtained in this paper on our finer lattice, $g_c=0.71(7)$. 
\item With the experimental $g_c$ and by using the isospin symmetry  we can evaluate the width of the $D^{\ast 0}$-meson as
\bea
\Gamma(D^{\ast 0})=  \Gamma(D^{\ast 0} \to D^- \pi^+) + \Gamma(D^{\ast 0} \to D^0 \pi^0) + \Gamma(D^{\ast 0} \to D^0 \gamma).
\eea
The first  channel turns out to be kinematically forbidden. For the last one, instead, we have the experimental information that  $B(D^{\ast 0}\to D^0\gamma) = 0.381(29)$~\cite{PDG}, so that with the $g_c$ value 
extracted from eq.~(\ref{g-cleo}) we have
\bea\label{lifetime0}
\Gamma(D^{\ast 0})= {1\over \, 1-(0.381\pm 0.029)\,  }\,  {  m_{D^0}  |\vec k_{\pi^0}^\prime|^3\over 12 \pi\ m_{D^{\ast 0}}\ f_\pi^2 } \, g_c^2 \, =\,  68\pm 17\ \kev\,,
\eea
where $|\vec k_{\pi^0}^\prime|= 43$~MeV. The above result  is well below an old --but the only available-- experimental bound,  $ \Gamma (D^{\ast 0}) < 2.1\ \mev$~\cite{abachi}.  
\item To be able to compare our results for $g_{D^\ast D\gamma}$ couplings with experiment, we first combine $\Gamma(D^{\ast +})=96\pm 22$~keV~\cite{cleo}, and $B(D^{\ast +}\to D^+\gamma) = 0.016(4)$~\cite{PDG}, 
to get
\bea
B(D^{\ast +}\to D^+\gamma) \times \Gamma(D^{\ast +}) = {\alpha_{\rm em}\over 3} g_{D^\ast D^+\gamma}^2\, k_{\gamma}^3 \ \Rightarrow\, g_{D^{\ast +} D^+\gamma}= 0.50(8)\ \gev^{-1}\,,
\eea
where, in this case,  $k_{\gamma}=135.6$~MeV. Similarly, from  the result in eq.~(\ref{lifetime0})  and $B(D^{\ast 0}\to D^0\gamma) = 0.381(29)$~\cite{PDG}, we obtain
\bea
g_{D^{\ast 0} D^0\gamma}= 2.02(26)\ \gev^{-1}\,.
\eea
We see that these results are in a quite  good agreement with values obtained on our finer lattice, namely  $g_{D^{\ast +} D^+\gamma}= -0.2(3)~\gev^{-1}$, and 
$g_{D^{\ast 0} D^0\gamma}=2.0(6)~\gev^{-1}$.
\item 
We can also compare to the experimentally measured ratio~\cite{PDG, aubert} 
\bea
R_0={\Gamma(D^{\ast 0}\to D^0\pi^0)\over \Gamma(D^{\ast 0}\to D^0\gamma )} = 1.74\pm 0.02\pm 0.13\,.
\eea
From our data we have $R_0 = 1.1^{+1.0}_{-0.7}$ at $\beta=5.29$, and $R_0=2.4^{+1.8}_{-1.3}$ at $\beta=5.40$. 
Broadly speaking both our lattice results are consistent with experiment. 
It will be interesting if one can increase the accuracy by which this ratio is computed and either check on the validity of  various lattice actions that are currently used in the literature, or 
to check on the chiral extrapolations.
\end{itemize}
The ratio $R_0$, as well as the quantities discussed in this section, are interesting because they do not  involve the weak interactions (CKM matrix elements) 
and the comparison between lattice QCD and experiment requires less assumptions.

Finally, since our result for the soft photon coupling is new, we should also compare it to the existing results in the literature. In table~\ref{table:compare} we collect the results obtained 
by using various quark models and various QCD sum rule techniques and compare them to the values obtained in this paper. In extracting the couplings 
$g_{D^\ast D \gamma}$ from various papers, we used  eq.~(\ref{eq:2}) and the results for the decay widths quoted in the cited papers.
\begin{table}[t!]
\hspace*{-5.5mm}
{\scalebox{.92}{ \begin{tabular}{|c|c|c|c|c|c|}
\hline
{\phantom{\Huge{l}}}\raisebox{-.1cm}{\phantom{\Huge{j}}}
Method & Ref. & $\Gamma(D^{\ast +}\to D^+\gamma)$&  {$\Gamma(D^{\ast 0}\to D^0\gamma)$} &$g_{D^{\ast +} D^+\gamma}$& $g_{D^{\ast 0} D^0\gamma}$  \\
 \hline
{\phantom{\Huge{l}}}\raisebox{-.1cm}{\phantom{\Huge{j}}}
\hspace*{-1cm}{\small\sl Light Front Model} & \cite{choi}  & $0.90(2)$  & $20.0(3)$ &  $0.4$  &  $1.8$\\ 
{\phantom{\Huge{l}}}\raisebox{-.1cm}{\phantom{\Huge{j}}}
  & \cite{jaus}  & $0.56$  & $21.7$ &  $0.3$  &  $1.9$\\ 
{\phantom{\Huge{l}}}\raisebox{-.1cm}{\phantom{\Huge{j}}}
\hspace*{-1cm}{\small \sl Chiral Quark Model}&\cite{goity}  & $1.0\div 1.5$  & $38.5\div 43.5$ &  $0.4\div 0.5$  &  $2.5\div 2.6$\\ 
{\phantom{\Huge{l}}}\raisebox{-.1cm}{\phantom{\Huge{j}}}
& \cite{chiral-model-2}  & $0.15\div 0.25$  & $11\div 13$ &  $0.16\div 0.20$  &  $1.4\div 1.5$\\ 
{\phantom{\Huge{l}}}\raisebox{-.1cm}{\phantom{\Huge{j}}}
\hspace*{-.7cm}{\small\sl Schr\"odinger-like Model}(``S")& \cite{galkin}  & $0.28$   & $17.4$ &  $0.2$  &  $1.7$\\ 
\hspace*{3.75cm}(``V")  &  & $0.6$   & $14.3$ &  $0.3$  &  $1.5$\\ 
{\phantom{\Huge{l}}}\raisebox{-.1cm}{\phantom{\Huge{j}}}
\hspace*{-1cm}{\small\sl Salpeter-like Model} & \cite{colangelo}  & $0.46$  & $20.8$ &  $0.3$  &  $1.8$\\ 
{\phantom{\Huge{l}}}\raisebox{-.1cm}{\phantom{\Huge{j}}}
\hspace*{-1cm}{\small\sl Bag Model} & \cite{orsland}   & $1.72$  & $7.18$ &  $0.5$  &  $1.1$\\ 
\hline 
{\phantom{\Huge{l}}}\raisebox{-.1cm}{\phantom{\Huge{j}}}
\hspace*{-1cm}{\small \sl Moment $3^{pts}$-Sum Rule} & \cite{dosch}   & $0.03(8)$  & $7.3(2.7)$ &  $0.02(11)$  &  $1.1(2)$\\ 
{\phantom{\Huge{l}}}\raisebox{-.1cm}{\phantom{\Huge{j}}}
\hspace*{-.65cm}{\small\sl Light Cone Sum Rule (LCSR)} & \cite{aliev}  & $1.5$  & $14.4$ &  $0.5$  &  $1.5$\\ 
\hline 
{\phantom{\Huge{l}}}\raisebox{-.1cm}{\phantom{\Huge{j}}}
\hspace*{-.65cm}{\small\sl Lattice QCD} & this work  & $0.8(7)$  & $27(14)$ &  $-0.2(3)$  &  $2.0(6)$\\ 
\hline 
{\phantom{\Huge{l}}}\raisebox{-.1cm}{\phantom{\Huge{j}}}
\hspace*{-.65cm}{\small\sl Experiment} & \cite{PDG}  & $1.54(52)$  & $26(7)$ &  $0.50(8)$  &  $2.02(26)$\\ 
\hline 
\end{tabular} }}
\caption{\label{table:compare}
\footnotesize  The table of results for the radiative $D^\ast$ meson decays as computed by various quark models, QCD sum rules,  lattice QCD and those extracted from experiment in the way discussed in Sec.~5 of 
this paper. ``S" /``V" labels the scalar/vector potential in the model of ref.~\cite{galkin}.}
\vspace*{-.3cm}
\end{table}
To that list we should add ref.~\cite{Rohrwild} in which the radiative $D^\ast$-meson decay was used to fix the unknown parameter  $\chi(1.3~\gev ) \varphi(1/2)$, the product of susceptibility 
of the quark condensate and the twist-2 photon distribution amplitude at the middle point.  More precisely, the problem of  large contributions due to coupling to the first radial  excitations~\cite{us3} was 
circumvented by applying the QCD sum rule to the branching fraction, rather than the decay width alone.

\section{Summary and perspectives}

In this paper we report the results of the first lattice QCD study  of the $g_{D^\ast D\gamma}$ coupling, i.e. of the magnetic moment of the charmed vector meson. 
Our result for the neutral $D^\ast$ meson is fully consistent with the current experimental decay width $\Gamma(D^{\ast }\to D^+\gamma)$. 
On the other hand our value for the charged meson suffers from large statistical errors but it is small and both consistent with zero and with 
the measured $\Gamma(D^{\ast }\to D^+\gamma)$.

We also provide the results of our new computation of the axial coupling $g_{D^\ast D\pi}$, which corroborates the previous (quenched) result obtained on the lattice. 
Our value agrees with the experimental result for  $\Gamma(D^{\ast +})^{\rm CLEO}$, with a tendency towards its larger values. 

The results presented in this paper are obtained by using the non-perturbatively ${\cal O}(a)$-improved Wilson quarks, both dynamical and valence ones. 
In spite of the improved technique to extract the form factors using the twisted boundary conditions and the convenient double ratios, the gain is not that obvious 
because our statistical errors are large. 
Besides an obvious goal to reduce the statistical uncertainties one should also try and probe the smaller light quark masses and thereby arrive at the   
precision determination of the couplings $g_{D^\ast D\pi}$ and $g_{D^\ast D\gamma}$. We argued that the comparison of the accurate lattice results for these  quantities 
may be a good testing ground for various lattice QCD approaches (different quark actions, effective treatments of heavy quark, chiral extrapolations) against experiment, 
because they involve the calculation of hadronic matrix element but, being the strong and electromagnetic processes, no CKM parameter is needed. 
The approach used here could also be used to compute the $g_{K^\ast K\gamma}$ coupling for which the experimental value is already known to a $5\%$ accuracy.
 
Finally the unquenching made here is done with $N_{\rm f}=2$ dynamical quarks and a natural step forward would be to include the effects of 
the `strange' dynamical quark, $N_{\rm f}=2 \to 2+1$.

\section*{Acknowledgements}
We thank  the QCDSF collaboration for letting us use their gauge field configurations,   the {\it Centre de Calcul de l'IN2P3 \`a Lyon}, for giving us access to their computing facilities
and the partial support of `Flavianet' ,  EU contract MTRN-CT-2006-035482.

\end{document}